\documentclass[%
 reprint,
superscriptaddress,
nofootinbib,
 amsmath,amssymb,
prc,
floatfix,
]{revtex4-2}

\usepackage{graphicx}
\usepackage{dcolumn}
\usepackage{bm}
\usepackage{hyperref}
\hypersetup{pdftex,colorlinks,linkcolor = blue,urlcolor  = blue,citecolor = blue}

\usepackage{siunitx}
\usepackage{upgreek}
\DeclareSIUnit\noop{\relax}

\begin{document}

\preprint{APS/123-QED}

\title{Experimentally constrained $^{165,166}\text{Ho}(n,\gamma)$ rates and implications for the $s$ process}

\author{F.~Pogliano}
\email{francesco.pogliano@fys.uio.no}
\affiliation{Department of Physics, University of Oslo, N-0316 Oslo, Norway}

\author{A.~C.~Larsen}
\email{a.c.larsen@fys.uio.no}
\affiliation{Department of Physics, University of Oslo, N-0316 Oslo, Norway}

\author{S.~Goriely}
\affiliation{Institut d’Astronomie et d’Astrophysique, Université Libre de Bruxelles, CP 226, B-1050 Brussels, Belgium}

\author{L.~Siess}
\affiliation{Institut d’Astronomie et d’Astrophysique, Université Libre de Bruxelles, CP 226, B-1050 Brussels, Belgium}

\author{M.~Markova}
\affiliation{Department of Physics, University of Oslo, N-0316 Oslo, Norway}

\author{A.~G\"{o}rgen}
\affiliation{Department of Physics, University of Oslo, N-0316 Oslo, Norway}

\author{J.~Heines}
\affiliation{Department of Physics, University of Oslo, N-0316 Oslo, Norway}

\author{V.~W.~Ingeberg}
\affiliation{Department of Physics, University of Oslo, N-0316 Oslo, Norway}

\author{R.~G.~Kjus}
\affiliation{Department of Physics, University of Oslo, N-0316 Oslo, Norway}

\author{J.~E.~L.~Larsson}
\affiliation{Department of Physics, University of Oslo, N-0316 Oslo, Norway}

\author{K.~C.~W.~Li}
\affiliation{Department of Physics, University of Oslo, N-0316 Oslo, Norway}

\author{E.~M.~Martinsen}
\affiliation{Department of Physics, University of Oslo, N-0316 Oslo, Norway}

\author{G.~J.~Owens-Fryar}
\affiliation{Department of Physics and Astronomy, Michigan State University, East Lansing, MI 48824, USA}
\affiliation{Facility for Rare Isotope Beams, Michigan State University, East Lansing, MI 48824, USA}

\author{L.~G.~Pedersen}
\affiliation{Department of Physics, University of Oslo, N-0316 Oslo, Norway}

\author{S.~Siem}
\affiliation{Department of Physics, University of Oslo, N-0316 Oslo, Norway}

\author{G.~S.~Torvund}
\affiliation{Department of Physics, University of Oslo, N-0316 Oslo, Norway}

\author{A.~Tsantiri}
\affiliation{Department of Physics and Astronomy, Michigan State University, East Lansing, MI 48824, USA}
\affiliation{Facility for Rare Isotope Beams, Michigan State University, East Lansing, MI 48824, USA}

\date{\today}

\begin{abstract}
The $\gamma$-ray strength function and the nuclear level density of $^{167}$Ho have been extracted using the Oslo method from a $^{164}\text{Dy}(\alpha,p\gamma)^{167}$Ho experiment carried out at the Oslo Cyclotron Laboratory. 
The level density displays a shape that is compatible with
the constant temperature model in the quasicontinuum, while the strength function shows structures indicating the presence of both a scissors and a pygmy dipole resonance.
Using our present results as well as data from a previous $^{163}\text{Dy}(\alpha,p\gamma)^{166}$Ho experiment, the $^{165}\text{Ho}(n,\gamma)$ and $^{166}\text{Ho}(n,\gamma)$ MACS uncertainties have been constrained. 
The possible influence of the low-lying, long-lived 6~keV isomer $^{166}$Ho in the $s$ process is investigated in the context of a 2~$M_\odot$, [Fe/H]=-0.5 AGB star. 
We show that the newly obtained $^{165}\text{Ho}(n,\gamma)$ MACS affects the final $^{165}$Ho abundance, while the $^{166}\text{Ho}(n,\gamma)$ MACS only impacts the enrichment of $^{166,167}$Er to a limited degree due to the relatively rapid $\beta$ decay of the thermalized $^{166}$Ho at typical $s$-process temperatures.
\end{abstract}

\maketitle



\section{Introduction and motivation}

The two main mechanisms responsible for the creation of elements heavier than iron in the universe are the $s$ and the $r$ processes, standing for slow and rapid neutron-capture process, respectively~\cite{B2HF, Cameron57}. 
The $r$ process lasts for a few seconds 
and involves neutron densities of $N_n \gtrsim 10^{20}$ cm$^{-3}$~(see, e.g., Refs.~\cite{ArnouldGoriely2007}). 
Such extremely high neutron densities will create very exotic, neutron-rich nuclei close to the neutron drip line, and will eventually $\beta$-decay to stability when the neutron flux is exhausted.

In contrast, the $s$ process  involves neutron densities of $N_n \leq 10^{10}$ cm$^{-3}$ and may last for thousands of years during the asymptotic giant branch (AGB) phase of low-mass stars~\cite{kappeler2011}. 
At these low neutron densities, neutron captures usually take place on stable or very long-lived nuclei, as the neutron-capture timescale is longer than the one for the $\beta$-decay for most of the unstable nuclei.
This means that $s$ process nucleosynthesis follows a relatively narrow path along the valley of $\beta$-stability up to Pb and Bi.
However, some $\beta$-unstable neutron-rich nuclei have longer lifetimes than others, and if their lifetimes are comparable to the average timescale for neutron capture, they become so-called branching points along the $s$ process path. 
In these cases, astrophysical conditions such as neutron density and temperature may influence the specific path the $s$ process takes, and a precise knowledge of the nuclear properties of the involved nuclei is paramount for the correct description of the nucleosynthesis flow~\cite{kappeler2011}.
Examples of $s$-process branching points include $^{85}\text{Kr}$ and $^{151}\text{Sm}$, where their location in the nuclear chart in between two stable nuclei gives separable branches that the reaction flow may follow (see  Ref.~\cite{kappeler2011} and references therein). 

One case of interest is the odd-odd $^{166}$Ho. As $^{165}$Ho is the only stable isotope of this element, $^{166}$Ho is made during the $s$ process.
Although its ground state $\beta$-decays to $^{166}$Er rather fast ($T_{1/2} \approx 26$h), $^{166}$Ho has a very low-lying ($E_x \approx 6$ keV) $7^-$ isomeric state that also $\beta$-decays to $^{166}$Er, but with a much longer half-life of about 1200 years~\cite{Faler1965}.
This half-life is on the same timescale as the $s$ process, which means that this branching could affect the final abundance of $^{165}$Ho as well as the isotopic abundance ratio of $^{166}$Er/$^{167}$Er.

Assuming $^{166}$Ho to be thermalized under typical $s$-process conditions (which should be a valid assumption according to Misch \textit{et al.}~\cite{Misch21}), the correct estimate of its impact requires knowledge of various nuclear properties, such as the $^{165}$Ho($n,\gamma$) reaction rate, the $^{166}$Ho($n,\gamma$) reaction rate, and the $^{166}$Ho $\beta$-decay rate. 
While the latter has been estimated by  \citet{Takahashi87}, the $^{165}$Ho($n,\gamma$) cross section has been measured directly \cite{Bao2000}. 
In addition, both neutron-capture rates can be indirectly derived from experimentally extracted nuclear level densities and $\gamma$-strength functions for $^{166}$Ho and $^{167}$Ho using the Hauser-Feshbach formalism~\cite{Hauser1952,TALYS196,RAUSCHERTHIELEMANN}.

In this work, we aim at clarifying the impact of $^{166}$Ho on the $s$ process by using experimentally constrained $^{165}$Ho($n,\gamma$) and $^{166}$Ho($n,\gamma$) rates in $s$-process simulations. 
In Sec.~\ref{secExperimental_setup} we present the results of the $^{164}$Dy($\alpha,p\gamma$)$^{167}$Ho experiment carried out at the Oslo Cyclotron Laboratory. 
Using the Oslo method, we are able to extract the level density and $\gamma$ strength function, which are used as input to
calculate the $^{166}$Ho($n,\gamma$) Maxwellian-averaged cross sections as described in Sec.~\ref{secNCrates}. 
In Sec.~\ref{secAstrophysics}, $s$-process calculations in AGB stars are performed, and the impact of the newly derived neutron-capture rates on the final abundances is discussed.

\section{Extraction of the nuclear level density and the \ensuremath{\gamma}-ray strength function}\label{secExperimental_setup}

While nuclear energy levels and reduced transition probabilities can be measured within the discrete region using spectroscopy methods, this task becomes increasingly difficult when going higher up in excitation energy. 
Here, levels become so close to each other that it is very difficult to distinguish them experimentally. 
When the mean level spacing $D$ becomes so small that $D^{-1} \geq 50-200$~MeV$^{-1}$, the nuclear properties are better described in terms of average statistical quantities: the nuclear level density (NLD) and the $\gamma$-ray strength function (GSF). 
These two quantities, apart from being essential ingredients to calculate neutron-capture rates within the Hauser-Feschbach framework~\cite{Hauser1952}, they may also reveal collective effects in the nucleus of interest for nuclear structure.
The total NLD for all spins and both parities is usually written as $\rho(E_x)$ and gives information on the number of energy levels per excitation-energy bin.
The GSF, written as $f^{XL}$, gives  information on the electromagnetic response of the nucleus and the probabilities for $\gamma$-decay of electric or magnetic character $X$ and multipolarity $L$.
The GSF is defined as~\cite{Bartholomew1973}
\begin{equation}\label{eq:Bartholomew73}
    f^{XL}(E_x,E_\gamma,J,\pi) = \frac{\langle\Gamma_\gamma^{XL}(E_x,E_\gamma,J,\pi)\rangle}{D(E_x,E_\gamma,J,\pi)E_\gamma^{2L+1}},
\end{equation}
where $E_x$ is the initial excitation energy, $E_\gamma$ is the transition energy, $J$ is the  angular momentum, $\pi$ is the parity, $\left< \Gamma_\gamma^{XL} \right>$ is the average partial $\gamma$-decay width, and $D$ is the mean level spacing for the specific class of quantum levels considered in the average. 
The partial widht $\left< \Gamma_\gamma^{XL} \right>$ can be related to the transmission coefficient $\mathcal{T}^{XL}$ by~\cite{BlattWeisskopf}
\begin{equation}\label{eq:GXLdefinition}
    \langle\Gamma_\gamma^{XL}(E_x,E_\gamma,J,\pi)\rangle = \mathcal{T}^{XL}(E_x,E_\gamma,J,\pi)\frac{D(E_x,E_\gamma,J,\pi)}{2\pi}.
\end{equation}
By joining Eqs.~(\ref{eq:Bartholomew73}) and (\ref{eq:GXLdefinition}), the transmission coefficient $\mathcal{T}^{XL}$ and the GSF $f^{XL}$ can be related through
\begin{equation}\label{Tgeneral}
    f^{XL}(E_x,E_\gamma,J,\pi) = \frac{\mathcal{T}^{XL}(E_x,E_\gamma,J,\pi)}{2\pi E_\gamma^{2L+1}}.
\end{equation}
Here, $E_x$, $J$ and $\pi$ may be averaged out using the generalized Brink-Axel hypothesis~\cite{Brink, Axel}, shown to hold for Dy nuclei~\cite{Renstroem2018}, and it is usually sufficient to consider dipole radiations $E1$ and $M1$ that dominate in the quasicontinuum region (see, e.g. Ref.~\cite{KopeckyUhl1990}). 
These two assumptions simplify Eq.~(\ref{Tgeneral}) to
\begin{equation}\label{Tgeneral2}
    f(E_\gamma) = \frac{\mathcal{T}(E_\gamma)}{2\pi E_\gamma^3}.
\end{equation}

The NLD and the GSF can be extracted from experimental data using the Oslo method. In the following we go through the experimental setup, the experiment itself and a brief description of the data analysis method.

\subsection{Experimental setup}

The experiment was carried out at the Oslo Cyclotron Laboratory in October 2022 and aimed at measuring $p-\gamma$ coincidences from the $^{164}\text{Dy}(\alpha,p\gamma)^{167}\text{Ho}$ reaction. 
A beam of $\alpha$ particles with $\approx 1.3$~nA intensity was accelerated to 26~MeV by the MC-35 Scanditronix cyclotron, and the beam impinged on a $^{164}$Dy self-supporting target, $1.73$~mg/cm$^2$ thick and with 98.5\% enrichment.
The target was placed in the center of the Oslo SCintillator ARray (OSCAR) and the Silicon Ring (SiRi) detector arrays, which recorded particle-$\gamma$ coincidences. OSCAR~\cite{Vetle_in_preparation, zeiser2021} is an array of 30 cylindrical (3.5''$\times$8.0'') LaBr$_3$(Ce) scintillator detectors mounted on a truncated icosahedron frame, with an energy resolution of 2.7\% full-width half maximum at $E_\gamma = 662$~keV and a prompt timing peak with time resolution of $\approx 1.8$~ns (standard deviation) for this experiment. 
SiRi~\cite{Guttormsen_NIMA_2011} is a $\Delta E$-$E$ particle telescope array, which consists of eight silicon-telescope modules in a ring configuration covering $126^\circ-140^\circ$ (corresponding to about 6\% of $4\pi$) in backward angles. 
Each module consists of a thick (1550~$\upmu$m) $E$ back detector and a thin (130~$\upmu$m) $\Delta E$  front detector. 
Each front detector is segmented in eight strips covering about $2^\circ$ each, while the back detector is not segmented.
The different energies deposited in the $E$ and $\Delta E$ detectors allow us to discriminate between different ejectiles, so that the data from the $(\alpha,p)$ channel could be selected. 
The specific reaction kinematics allows us to calculate the excitation energy $E_x$ the residual nucleus is left in, and associate this to its corresponding $\gamma$ spectrum. 
By plotting the detected $\gamma$ rays  against $E_x$ we obtain a matrix called the coincidence matrix, which is the starting point for extracting the NLD and the GSF using the Oslo method.

\subsection{The Oslo method and normalization details}
\begin{table*}[ht]
\caption{\label{tab:NLD_params} Parameters used for the $^{167}$Ho NLD and GSF normalization.  
The parameters $E_0$ and $T_\textrm{CT}$ are determined from a fit to our data points at high $E_x$ together with the calculated $\rho(S_n)$ value.  
The $D_0$ and $\langle \Gamma_\gamma \rangle$ values are taken from Ref.~\cite{MughabghabS.F.2018Aonr}, but with an uncertainty estimation as described in the text.
The two values for the spin-cutoff parameter $\sigma^2_I(S_n)$ are from the FG formula~\cite{GilbertCameron} and from the RMI formula~\cite{EgidyBucurescu2005, *EgidyBucurescu2006} (see text and Ref.~\cite{pogliano23}). 
}
\begin{ruledtabular}
\begin{tabular}{cccccccc}
$E_0$ (MeV) & $T_\textrm{CT}$ (MeV) & $D_0$ (eV) & $I_t$ & $\sigma^2_I$(FG) & $\sigma^2_I$(RMI) & $\sigma^2_d$ &  $\langle \Gamma_\gamma\rangle$ (meV)\\ \hline
$-1.836$     & 0.620 & $2.32(77)$ & 7 & 5.68 & 7.10 & 2.96 & $89(9)$  \\
\end{tabular}
\end{ruledtabular}
\end{table*}
The $\gamma$ rays measured with OSCAR will inevitably be convoluted with the detector response~\cite{zeiser2021}. 
The \textit{unfolding} procedure~\cite{GUTTORMSEN1996} helps us correct for this convolution, and obtain a $\gamma$ ray spectrum for the full-energy peaks only. 
From the unfolded spectra we can obtain the first-generation $\gamma$-rays using the weighted subtraction technique by Guttormsen \textit{et al.}~\cite{GUTTORMSEN1987}.

By inspection of the first-generation matrix, we may select the region coinciding with the quasi-continuum, in this case between $E_x = 4.5$ and $7.0$~MeV, and limiting the $\gamma$ rays to $E_\gamma > 1.2$~MeV. 
From Fermi's Golden rule~\cite{Dirac1927,Fermi1950}, we may express the $\gamma$-decay probability $P(E_\gamma,E_x)$ for a nucleus at excitation energy $E_x$ to emit a $\gamma$ ray of energy $E_\gamma$ as~\cite{SCHILLER2000}
\begin{equation}\label{PTrhoraw}
    P(E_\gamma,E_x) \propto \mathcal{T}(E_\gamma) \rho(E_x - E_\gamma).
\end{equation}
Using a global $\chi^2$ minimization technique described in Ref.~\cite{SCHILLER2000}, we are then able to extract the functional shape of the NLD and the GSF from the selected region in the first-generation matrix.
Thus we obtain the solutions
\begin{subequations}
\begin{align}\label{UnnormRhoFa}
    \tilde{\rho}(E_x - E_\gamma) &= Ae^{\alpha (E_x - E_\gamma)}{\rho}(E_x-E_\gamma),\\
    \tilde{\mathcal{T}}(E_\gamma) &= Be^{\alpha E_\gamma}{\mathcal{T}}(E_\gamma), \label{UnnormRhoFb}
\end{align}
\end{subequations}
where $A$, $B$ and $\alpha$ are free parameters, and any choice of them gives an equally good fit to the first-generation matrix. 
To determine these parameter, we must make use of the known discrete energy levels at low $E_x$ and the level density at the neutron separation energy $S_n$ for the level density, and the average total radiative width $\langle \Gamma_\gamma \rangle$ for the strength function~\cite{OsloSystematicErrors, *OsloSystematicErrors2}.

The discrete energy levels are readily available at NUDAT~\cite{NUDAT}, while the value of $\rho(S_n)$ can be calculated from the measured level spacing $D_0$ of \textit{s}-wave neutron resonances and the spin-cutoff parameter $\sigma_I^2$ at $S_n$ by~\cite{SCHILLER2000}
\begin{equation}\label{eq:SnfromD0}
    \rho(S_n) = \frac{2\sigma_I^2}{D_0 \left[(I_t + 1)e^{-(I_t+1)^2/2\sigma_I^2} + I_t e^{-I_t^2/2\sigma_I^2}\right]},
\end{equation}
where $I_t$ is the spin of the $A-1$ isotope that is the target in the neutron-resonance experiment.

As our level-density data points do not reach $\rho(S_n)$ due to the lower limit on $E_\gamma$, the data has to be extrapolated up to $S_n$. 
The choice of extrapolation function is usually not important given that the lower $E_\gamma$ is not too large. 
Typically the extrapolation is done with either the back-shifted Fermi gas model~\cite{GilbertCameron, VONEGIDY1988} (BSFG):
\begin{equation}\label{BSFG}
    \rho_{\textrm{FG}} (E_x) = \frac{\exp\left( 2\sqrt{aU}\right)}{12 \sqrt{2} a^{1/4} U^{5/4} \sigma_I} 
\end{equation}
or the constant-temperature model (CT,~\cite{Ericson,GilbertCameron}):
\begin{equation}\label{rho_CT}
    \rho_{\textrm{CT}}(E_x) = \frac{1}{T_{\textrm{CT}}}\exp\left(\frac{E_x-E_0}{T_{\textrm{CT}}}\right),
\end{equation}
where $U = E_x-E_1$. 
Here $a$, $E_1$, $E_0$ and $T_{\textrm{CT}}$ are fitting parameters.
For $^{167}$Ho, the CT model was observed to fit the data at higher $E_x$ better than the BSFG one, and the values for $E_0$ and $T_\textrm{CT}$ were found to be $-1.836$~MeV and $0.620$~MeV, respectively. 

The value of $D_0$ can be retrieved from the \textit{Atlas of neutron resonances}~\cite{MughabghabS.F.2018Aonr}, where a calculated value of $D_0=2.32$~eV is provided using the long-lived $7^-$ isomer as target for thermal neutron capture. 
The only unknown left to calculate the level density at $S_n$ using Eq.~\ref{eq:SnfromD0} is  the spin-cutoff parameter at $S_n$. 
For this reason the normalization procedure used for $^{167}$Ho closely resembles the one used for $^{166}$Ho in Ref.~\cite{pogliano23} except for the fact that the CT model was used instead of the BSFG.

The choice of $\sigma^2_I$ is model dependent, as there is no experimental data on the spin distribution for all accessible spins at $S_n$ for these nuclei. 
Two widely used models are the rigid body of inertia formula as applied by von Egidy and Bucurescu ~\cite{EgidyBucurescu2005, *EgidyBucurescu2006} (here labeled RMI), or the Gilbert and Cameron approach~\cite{GilbertCameron} (here labeled FG). 
We have no reason to prefer one against the other, so we let $\sigma^2_I$ vary  between the FG value of $\sigma_I = 5.68$ and the RMI value of $\sigma_I =7.10$. 
We assume that the error in $\rho(S_n)$ is evenly distributed between the two $\sigma^2_I$ values, and otherwise decided by the uncertainty associated to the $D_0$ value. 
Unfortunately, the \textit{Atlas of neutron resonances}~\cite{MughabghabS.F.2018Aonr} does not provide an uncertainty to its recommended value. 
Considering that there are  three measured neutron resonances, $N_r = 3$, the uncertainty was estimated to be 33\% using the $\Delta D_0/D_0=1/N_r$ formula from RIPL3~\cite{RIPL}. 
For the $\sigma^2_I$ dependence on excitation energy, we follow Refs.~\cite{Guttormsen2017,RIPL}  and assume $\sigma^2(E_x)$ to be linearly dependent with respect to the excitation energy:
\begin{equation}\label{eq:spin_cutoff_Alex}
\sigma^2(E_x) = \sigma_d^2 + \frac{E_x - E_d}{S_n - E_d}[\sigma^2_I - \sigma_d^2],
\end{equation}
where $\sigma_d^2$ is the spin-cutoff parameter at a low excitation energy $E_d$. In our case, $\sigma_d^2$ was found to be 2.96 at $E_d~=~0.220$~MeV.
The fit of the NLD to the discrete levels is done in a similar way as Refs.~\cite{Pogliano22, pogliano23}, where we chose the $E_x$ interval with the most complete level scheme.

\begin{figure}
\includegraphics[width=0.50\textwidth]{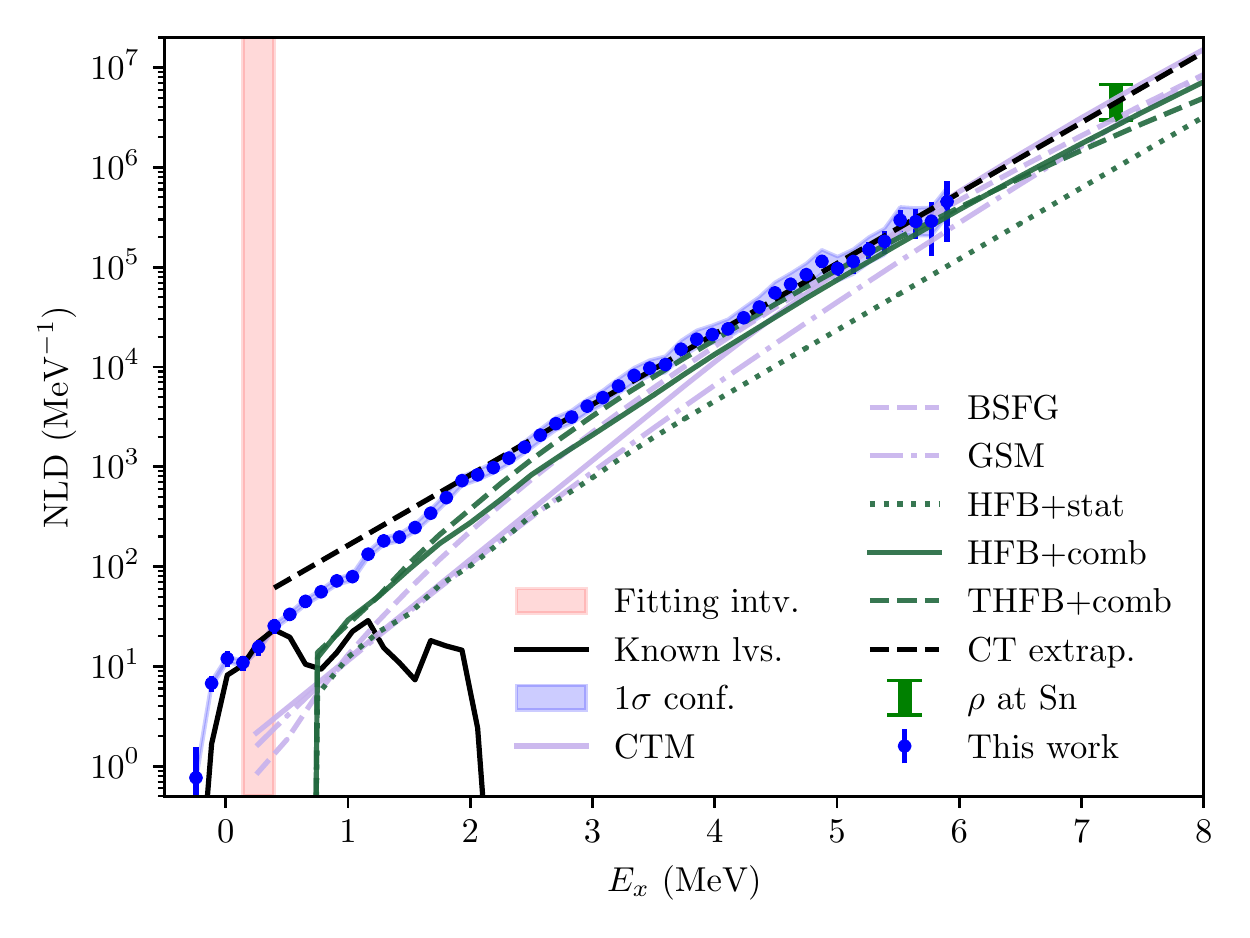}
\caption{\label{fig:nld} (Color online) The normalized NLD compared to the theoretical models used in TALYS 1.96~\cite{TALYS196}. For an overview of the models and references, see text.
The error bars indicate the statistical and systematic uncertainties from the Oslo method, and the uncertainty band shows the systematic errors from the normalization procedure. The vertical, pink-shaded band (light grey) indicates the region used for fitting the extracted NLD to the known levels of $^{167}$Ho from Ref.~\cite{NUDAT}.}
\end{figure}

\begin{figure}
\includegraphics[width=0.50\textwidth]{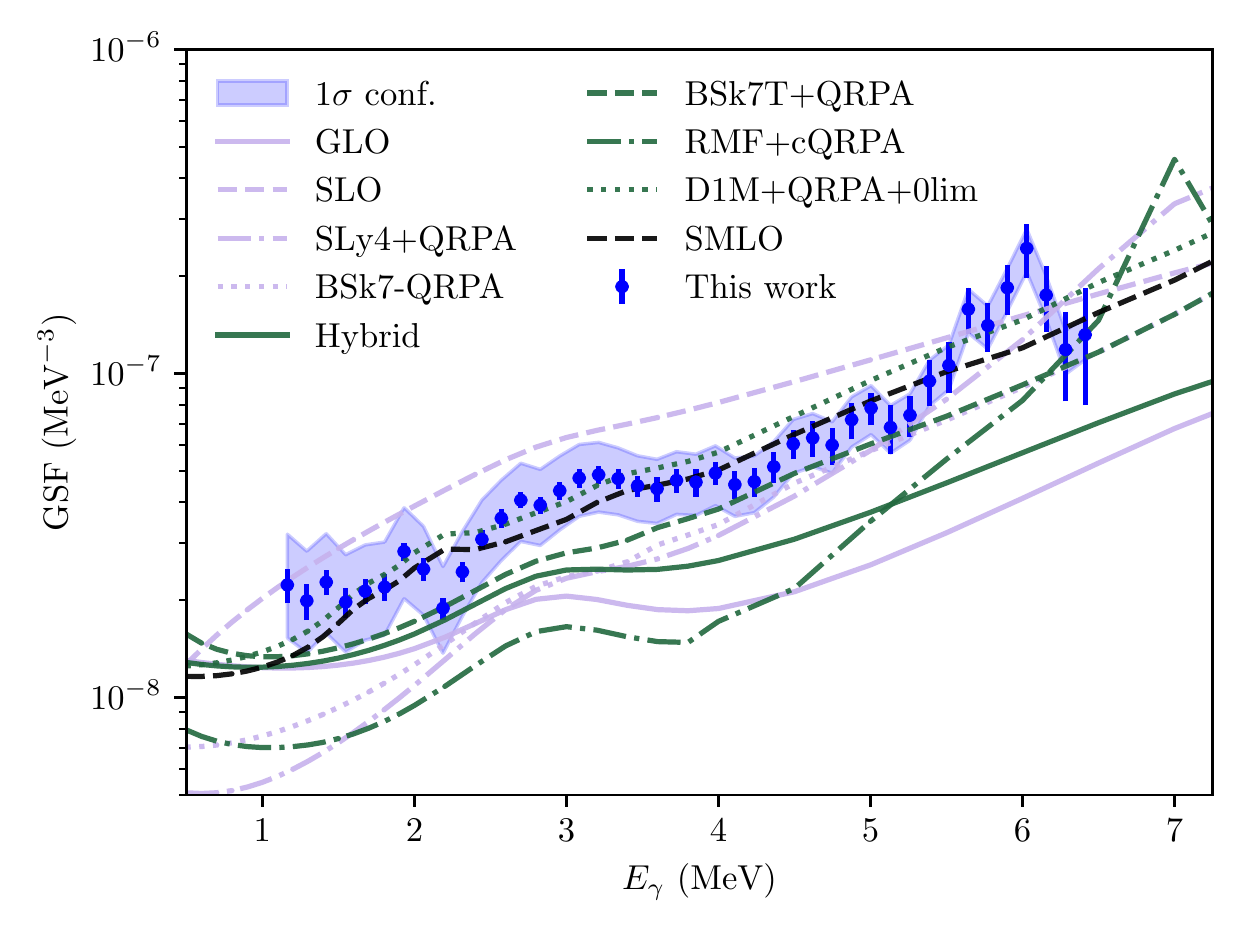}
\caption{\label{fig:gsf} (Color online) The normalized GSF compared to the theoretical models used in TALYS 1.96~\cite{TALYS196}. For all the $E1$ models listed in the plot, the $M1$ SMLO~\cite{Goriely18b} was added, except for the D1M+QRPA-0lim $E1$ model, where the corresponding D1M+QRPA-0lim $M1$ was used~\cite{Goriely2018}. See text for an overview of the models and their references.
Uncertainties are as in Fig.~\ref{fig:nld}.}
\end{figure}

For the normalization of the GSF we use the average, total radiative width $\langle \Gamma_\gamma \rangle$ \cite{OsloSystematicErrors, OsloSystematicErrors2}, also available in the \textit{Atlas of Neutron Resonances} where it is given as $88.5$~meV~\cite{MughabghabS.F.2018Aonr}. Again, the uncertainty in this quantity is not provided, but the value of $\langle \Gamma_\gamma \rangle$ is quite similar for neighboring rare-earth nuclei; from systematics of these $\langle \Gamma_\gamma \rangle$ values we estimate the uncertainty to be 10\%, therefore we have used $\langle \Gamma_\gamma \rangle = 89(9)$ meV. 
We use the same method as in Refs.~\cite{Pogliano22, pogliano23} in order to propagate the systematic and statistical uncertainties from the fitting parameters to the normalized GSF.
An overview of all the values used for the NLD and GSF normalization can be found in Table \ref{tab:NLD_params}, and the normalized NLD and GSF are displayed in Figs.~\ref{fig:nld} and \ref{fig:gsf} respectively.

\subsection{Level density and strength function}
The normalized level density is compared with TALYS~\cite{TALYS196} models in Fig.~\ref{fig:nld}. 
The models used in the comparison are:
\begin{itemize}
\item The constant-temperature plus Fermi gas model (CTM)~\cite{GilbertCameron}.
\item The Back-shifted Fermi gas model (BSFG)~\cite{GilbertCameron, VONEGIDY1988}.
\item The Generalised Superfluid model (GSM)~\cite{ignatyuk79, ignatyuk93}.
\item The Skyrme-Hartree-Fock-Bogolyubov plus statistical model (HFB+Stat), tables from Ref.~\cite{GORIELY2001311}.
\item The Hartree-Fock-Bogoluybov plus combinatorial model (HFB+comb), tables from Ref.~\cite{PhysRevC.78.064307}.
\item The temperature-dependent Gogny-Hartree-Fock-Bogolyubov model (THFB+comb)~\cite{PhysRevC.86.064317}.
\end{itemize}
In general, the models do not agree very well with the data at low excitation energies, but the agreement improves somewhat  for $E_x \geq 5$~MeV.

The comparison between the extracted GSF and the TALYS models can be seen in Fig.~\ref{fig:gsf}. 
The models here include both $E1$ and $M1$ radiation. The $E1$ models used are:
\begin{itemize}
\item The Kopecky-Uhl generalized Lorentzian (GLO)~\cite{KopeckyUhl1990}.
\item The Brink-Axel standard Lorentzian (SLO)~\cite{BRINK57,Axel}.
\item The Hartree-Fock-BCS plus QRPA tables based on the SLy4 interaction (SLy4+QRPA)~\cite{GorielyKhan2002}.
\item The HFB plus QRPA calculation based on the BSk7 interaction (BSk7+QRPA)~\cite{GorielyKhan2004}.
\item The hybrid model (Hybrid)~\cite{Goriely_hybrid_model}.
\item The BSk7+QRPA model with $T$-dependent width (BSk7T+QRPA)~\cite{GorielyKhan2004}.
\item The relativistic mean field plus continuum QRPA calculation with $T$-dependent width (RMF+cQRPA)~\cite{Daoutidis2012}.
\item The Gogny-HFB plus QRPA calculation complemented by low-energy enhancement (D1M+QRPA+0lim)~\cite{Goriely2018}.
\item The simplified modified Lorentzian (SMLO)~\cite{Goriely18b}.
\end{itemize}

For the $M1$ strength component, the default $M1$ SMLO model~\cite{Goriely18b} with upbend was used, except for the D1M+QRPA where the corresponding  $M1$ strength was used~\cite{Goriely2018}.

We note that most models predict a structure centered at $\approx 3$~MeV on the tail of the GEDR compatible with an $M1$ scissors resonance (SR)~\cite{Heyde2010}, although our experimental results do not match their predicted magnitude. 
The structure at $\approx 6$~MeV can be interpreted as the $E1$ pygmy dipole resonance~\cite{SAVRAN2013210,BRACCO2019360} (PDR), but here we should be careful as the poor statistics from the experiment leads to rather big statistical uncertainties. 
The D1M+QRPA model~\cite{Goriely2018} and the Simplified Modified Lorentzian \cite{Goriely18b} do the best job at predicting the GSF as they have the correct magnitude, although none of them reproduce the two observed resonance-like structures.

The Oslo method does not allow to distinguish between $E1$ and $M1$ radiation. 
Therefore, to extract e.g. the SR integrated strength, the GSF is modeled using empirical functions and data from neighboring nuclei. 
The giant dipole resonance (GEDR) is known to be of $E1$ character and expected to be double-peaked for a deformed nucleus~\cite{DIETRICH1988199,Harakeh}. 
We therefore model the GEDR with two Lorentzian-type functions using the generalized Lorentzian (GLO) function by Kopecky and Uhl~\cite{KopeckyUhl1990}:
\begin{multline}\label{eq:GLO}
    f^{\textrm{GLO}}(E_\gamma) = \sum_{i=1}^2 ~ \frac{\sigma_{0,i}\Gamma_{0,i}}{3\pi^2 \hbar^2 c^2}\times \\ \left(\frac{E_\gamma\Gamma_K(E_\gamma,T_f)}{\left(E^2_\gamma - E^2_{0,i}\right)^2 + E_\gamma^2 \Gamma^2_K} + 0.7\frac{\Gamma_K(0,T_f)}{E_{0,i}^3}\right),
\end{multline}
where
\begin{equation}
    \Gamma_K(E_\gamma,T_f) = \frac{\Gamma_{0,i}}{E_{0,i}^2}\left(E^2_\gamma + 4\pi^2T_f^2\right)
\end{equation}
and $\Gamma_{K,0}=\Gamma_K(0,T_f)$. Here $E_{0,i}$, $\Gamma_{0,i}$, $\sigma_{0,i}$ and $T_f$ are parameters represententing the energy centroid, the width, the peak cross section of each peak ($i = {1,2}$), and the temperature of the final levels, respectively.
The functions are fitted to the neighboring $^{165}$Ho GEDR data as this is the closest nucleus with experimental GEDR data available, measured by Berman \textit{et al.}~\cite{BermanHo} and Bergere \textit{et al.}~\cite{BergereHo}. 
However, we chose to apply the more recent re-analysis and evaluation of the two experiments from Varlamov \textit{et al.}~\cite{Varlamov19}.

\begin{figure}
\includegraphics[width=0.50\textwidth]{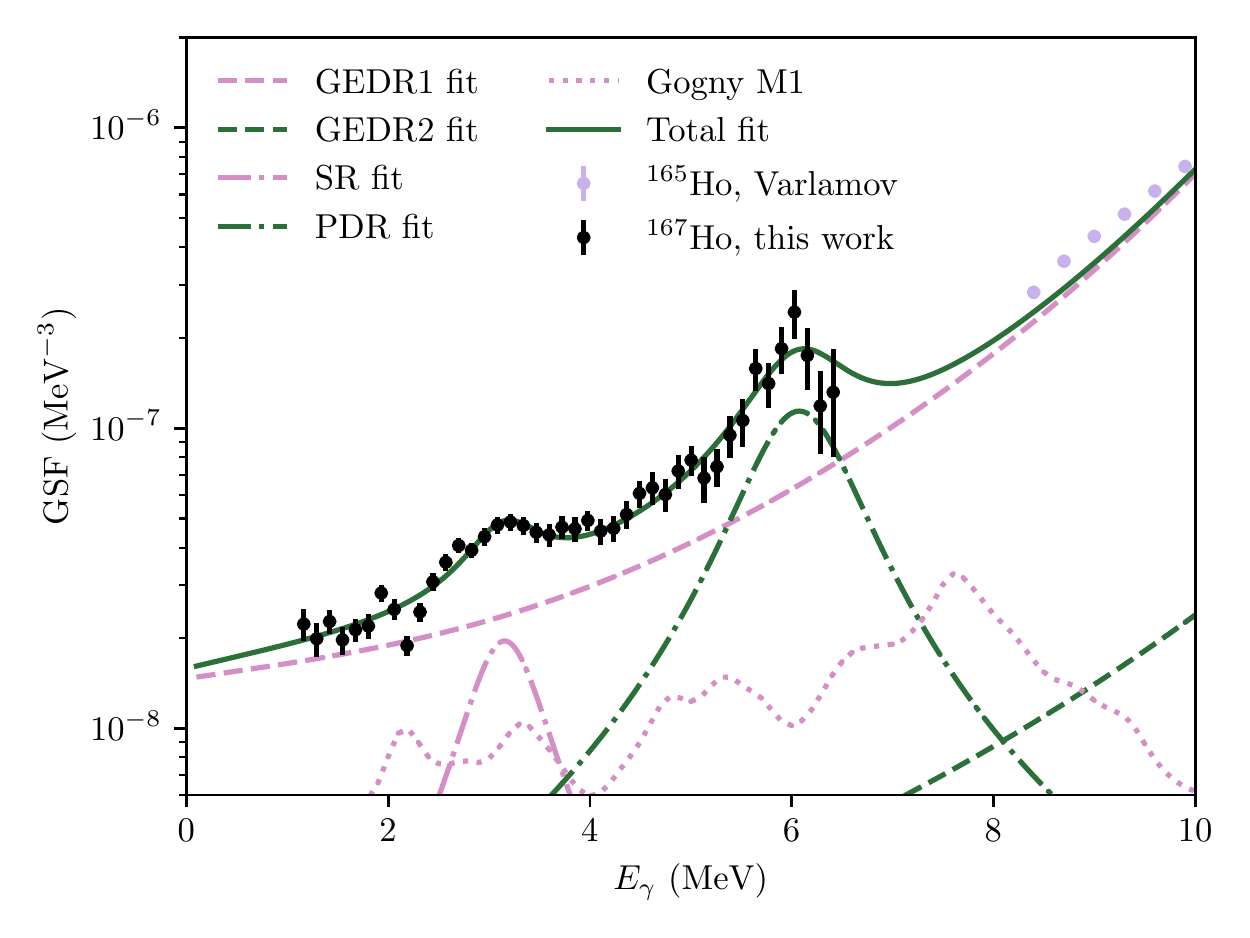}
\caption{\label{fig:gsf_fit} (Color online) The normalized GSF and the fit to the data using the emipirical functions described in the text. The $^{165}$Ho GEDR data from Varlamov \textit{et al.}~\cite{Varlamov19} is used fit the GEDR.  
The dotted line shows the theoretical QRPA predictions for the $M1$ GSF from Ref.~\cite{Gogny_QRPA}.}
\end{figure}

The PDR and the SR are fitted by a standard Lorentzian (SLO), defined as
\begin{equation}\label{eq:SLO}
    f^{\textrm{SLO}}(E_\gamma) = \frac{1}{3\pi^2 \hbar^2 c^2}\frac{\sigma_s\Gamma_s^2 E_\gamma}{\left(E^2_\gamma - E^2_s\right)^2 + E_\gamma^2 \Gamma^2_s},
\end{equation}
where $E_s$, $\Gamma_s$ and $\sigma_s$ are the  resonance parameters representing the energy centroid, the width and the peak cross section.
We see that the data is modeled relatively well, considering the above-mentioned big uncertainties concerning the pygmy-like structure at $E_\gamma \approx 6$~MeV.
The fit to the data allows for a clear separation between the contributions from the scissors and the pygmy-like structures from the  GEDR tail, and determining their respective strengths.
The spin-flip $M1$ resonance is also probably present, but its expected contribution centered around 8~MeV has likely a  magnitude far below the $E1$ contribution, so that we did not include it in the fit. 
All the fitting parameters are listed in Table~\ref{fittingparams}.

Of certain interest is the integrated \textit{upward} SR strength $B_\textrm{SR}$ that can be expressed as
\begin{equation}\label{eq:BSR}
    B_{\textrm{SR}}=\frac{(3\hbar c)^3}{16\pi}\int f_{\textrm{SR}}(E_\gamma)\textrm{d}E_\gamma,
\end{equation}
where $f_\textrm{SR}$ is the function describing the SR. 
This function could, for example, be the SLO function fitting the broad structure at $E_\gamma \approx 3$~MeV.
To obtain a lower bound for the integrated strength,  the $E1$ tail can be modeled as a simple exponential function going through two data points that are considered to be the $E_\gamma$ limits of the SR. 
Such an approach has been used by Agvaanluvsan \textit{et al.}~\cite{Agvaanluvsan2004}, Nyhus \textit{et al.}~\cite{Nyhus10}, Malatji \textit{et al.}~\cite{Malatji2021} and referred to as Method 3 in Pogliano \textit{et al.}~\cite{pogliano23}. 
By choosing $E_\gamma = 1.804$~MeV and 4.236~MeV as the limiting points, we calculate an integrated strength $B_\textrm{SR}=4.0(7)$$~$$\mu_N^2$. 
This is to be compared to the $B_\textrm{SR}$ calculated by integrating the SLO fit of the SR using the parameters in Table~\ref{fittingparams} between 0 and 10~MeV, giving a value of 6.3(10)~$\mu^2_N$. 
The latter value assumes the $E1$ contribution stemming from the GEDR tail to be smaller than for the former value, and is comparable to values obtained for $^{163,164}$Dy by Renstr{\o}m \textit{et al.}~\cite{Renstroem2018}.

\begin{figure}
\includegraphics[width=0.50\textwidth]{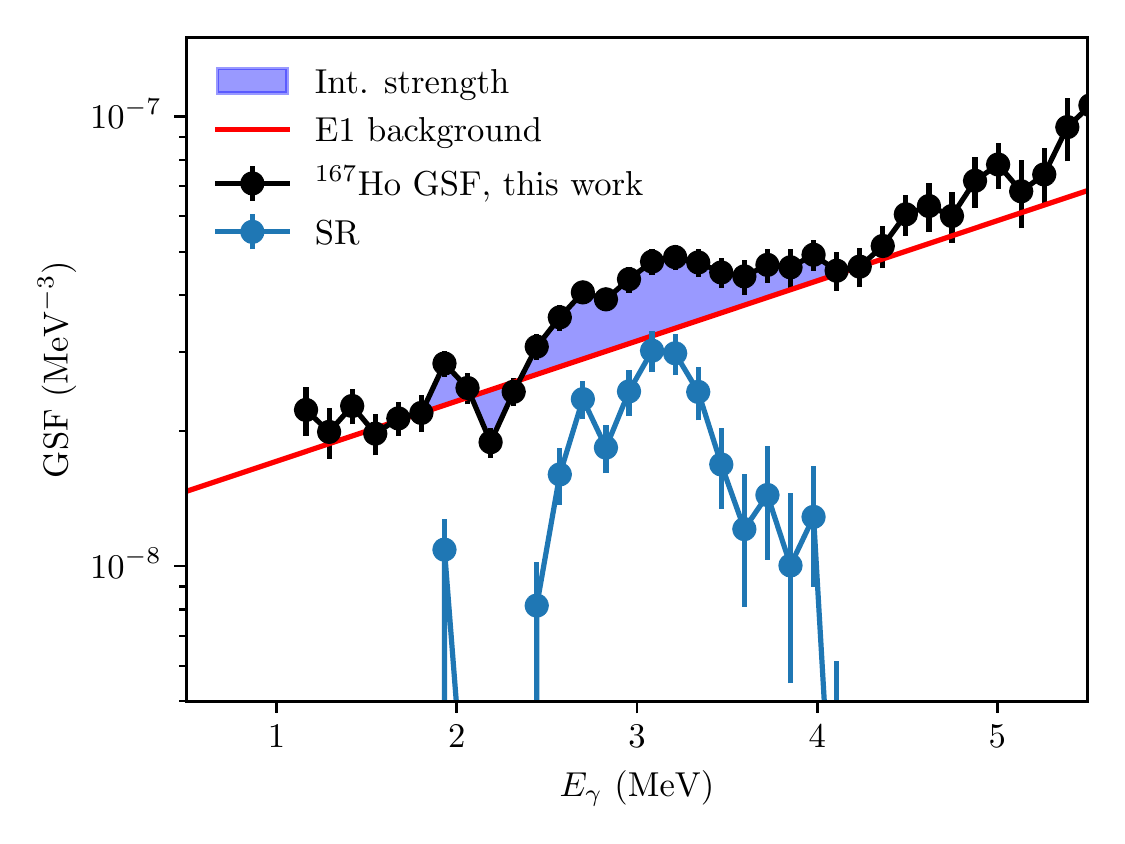}
\caption{\label{fig:Hildesmethod} (Color online) Estimation of the $E1$  strength using a simple exponential passing through two points of the GSF (see text). 
Data points from  this work are shown as black points, while the red line is the modeled $E1$ strength, and the blue area shows the residual strength. 
}
\end{figure}

\begin{table}
\caption{\label{fittingparams} 
The parameters used in the fit functions shown in Fig.~\ref{fig:gsf_fit}. 
}
\begin{ruledtabular}
\begin{tabular}{cccccc}
Function             & $T_f$ & $E_{0,s}$ & $\Gamma_{0,s}$ & $\sigma_{0,s}$ & $B_\textrm{SR}$ \\ 
                     & (MeV)   & (MeV) & (MeV) & (mb) & ($\mu^2_ I$) \\\hline

GEDR1                & 0.72(1) & 12.34(1) & 3.17(3) & 337(1) & - \\
GEDR2                & -         & 14.78(1) & 1.85(3) & 196(2)  & -\\
PDR                  & -         & 6.11(5)  & 1.20(3) & 8.0(6)  & -\\
SR$_\textrm{SLO}$    & -         & 3.19(5) & 0.87(11) & 0.72(7)  & 6.3(10)\\
SR$_\textrm{exp}$    & -         & -       & -       & -        &  4.0(7)\\
\end{tabular}
\end{ruledtabular}
\end{table}

\section{Neutron-capture rates}\label{secNCrates}

With our experimental data on the NLD and the GSF, and by using a neutron optical-model potential (OMP), we can employ the Hauser-Feshbach formalism~\cite{Hauser1952,RAUSCHERTHIELEMANN} in order to calculate the $(n,\gamma)$ cross section for the $N-1$ isotope. 
Here, we use the $^{166}$Ho data from Ref.~\cite{pogliano23}  and the present NLD and GSF of $^{167}$Ho from this work to calculate the $^{165}\text{Ho}(n,\gamma)$ and $^{166}\text{Ho}(n,\gamma)$ cross sections, respectively. 
The cross sections are calculated using the nuclear reaction code TALYS 1.96~\cite{TALYS196}. 
Information on the OMP cannot be extracted using the Oslo method, and we here rely on the OMP models implemented in TALYS. 
We use the phenomenological model by Koning and Delaroche~\cite{localomp} for both nuclei, where OMP parameters from experimental data are given for $^{165}\text{Ho}$, as well as the semi-microscopic Jeukenne-Lejeune-Mahaux (JLM) model by Bauge \textit{et al.}~\cite{JLMOMP}.

\begin{figure}
\includegraphics[width=0.50\textwidth]{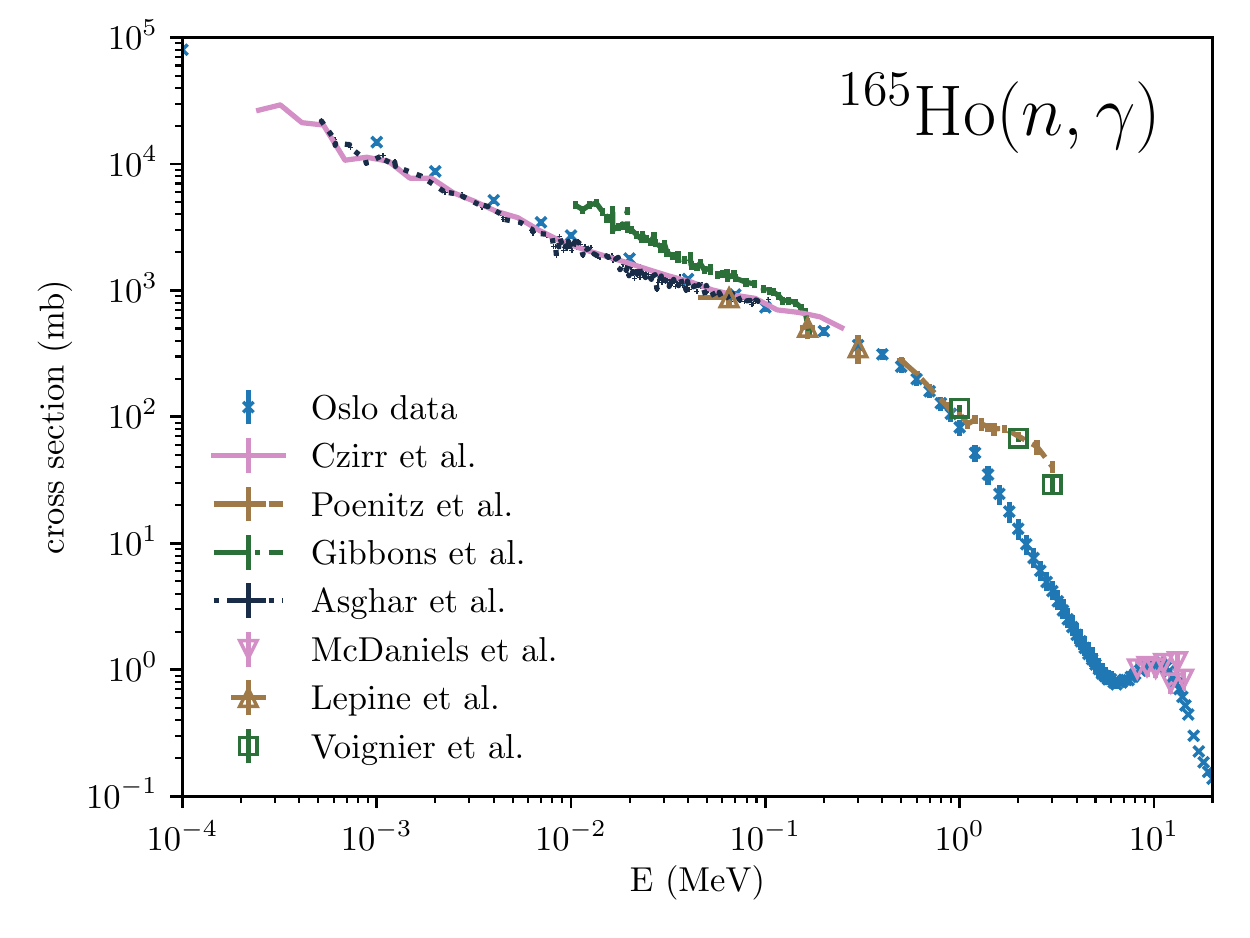}
\caption{\label{fig:directcapture} (Color online) The $^{165}\text{Ho}(n,\gamma)$ cross section calculated with the experimentally extracted NLD and GSF for $^{166}\text{Ho}$ compared to data from Czirr \textit{et al.}~\cite{Czirr73}, Poenitz \textit{et al.}~\cite{Poenitz1975}, Gibbons \textit{et al.}~\cite{Gibbons61}, McDaniels \textit{et al.}~\cite{McDaniels82}, Asghar \textit{et al.}~\cite{Asghar68}, Voignier \textit{et al.}~\cite{Voignier92} and Lepine \textit{et al.}~\cite{Lepine72}.}
\end{figure}

Our calculated $^{165}\text{Ho}(n,\gamma)$ cross section is compared with directly measured neutron-capture data from the literature in Fig.~\ref{fig:directcapture}. 
We observe that our experimentally-constrained cross section calculation agrees rather well with the data sets of Czirr \textit{et al.}~\cite{Czirr73}, Asghar \textit{et al.}~\cite{Asghar68}, Lepine \textit{et al.}~\cite{Lepine72} and McDaniels \textit{et al.}~\cite{McDaniels82}.
On the other hand, the other neutron-capture measurements seem to be significantly higher. 
The reason for this discrepancy in the directly measured cross sections as well as our result is not clear, and it would be desirable to perform new ($n,\gamma$) measurements on $^{165}$Ho to understand and resolve this issue.

The radiative neutron-capture cross section is a crucial ingredient to the neutron-capture rate $N_A\langle\sigma v\rangle(T)$ as seen from the reactivity equation (see,  e.g., Ref.~\cite{ArnouldGoriely2007})
\begin{multline}\label{ncrate}
    N_A\langle\sigma v\rangle(T)=\left(\frac{8}{\pi \tilde{m}}\right)^{1/2}\frac{N_A}{\left(k_B T\right)^{3/2}G_t(T)} \times \\\int_0^\infty \sum_\mu \frac{2J_t^\mu + 1}{2J_t^0 + 1}\sigma^\mu_{n\gamma}(E)E\exp\left[-\frac{E+E_x^\mu}{k_B T}\right]\textrm{d}E,
\end{multline}
where $N_A$ is Avogadro's number, $\tilde{m}$ the reduced target mass, $k_B$ is Boltzmann's constant, $T$ the temperature in the astrophysical environment, $J_t^0$ and $J_t^\mu$ are the ground state and the $\mu$th excited energy level spins respectively, $E_x^\mu$ the excitation energy of the $\mu$th energy level, $E$ the relative kinetic energy between the neutron and the target nucleus, $\sigma^\mu_{n\gamma}$ the $(n,\gamma)$ cross section for the target nucleus excited to the $\mu$'th state, and $G_t(T)$ the partition function given by
\begin{equation}
    G_t(T) = \sum_\mu \frac{2J_t^\mu + 1}{2J_t^0 + 1}\exp\left[\frac{-E_x^\mu}{k_B T}\right].
\end{equation}
From the radiative neutron-capture rate we can calculate the Maxwellian-averaged cross section $N_A\langle \sigma\rangle_T$ (MACS):
\begin{equation}\label{MACS}
    N_A\langle \sigma \rangle_T = \frac{N_A \langle \sigma v \rangle}{v_T},
\end{equation}
where $v_T = \sqrt{2k_B T/\tilde{m}}$ is the thermal velocity.

\begin{figure}
\includegraphics[width=0.50\textwidth]{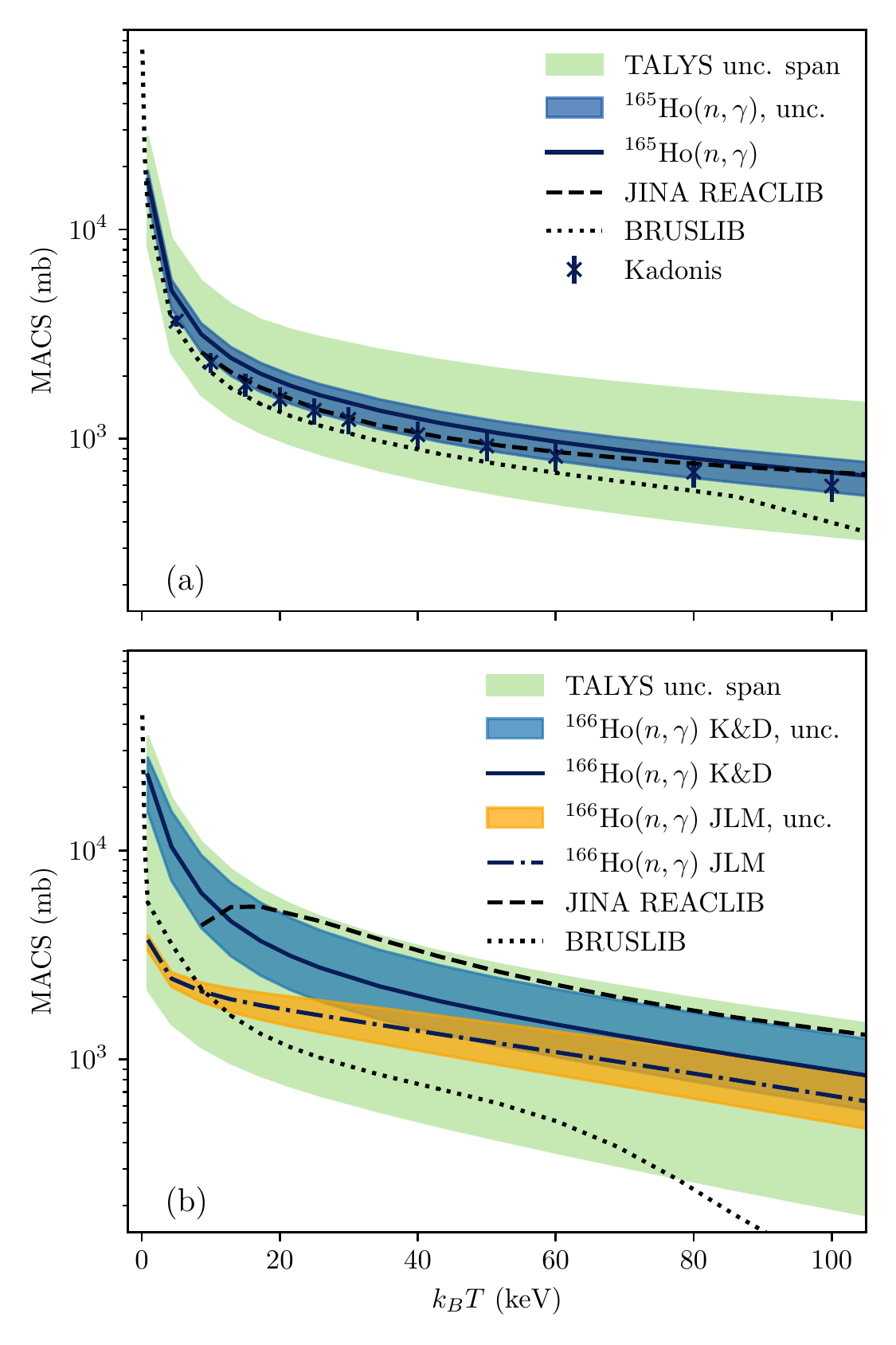}
\caption{\label{fig:MACSs} The Maxwellian-averaged cross sections for (a) $^{165}\text{Ho}(n,\gamma)$ and (b) $^{166}\text{Ho}(n,\gamma)$ using our data as input compared to the TALYS uncertainty range as well as JINA REACLIB~\cite{JINAREACLIB} and BRUSLIB~\cite{BRUSLIB} (see text). 
}
\end{figure}

The $^{165}\text{Ho}$ and $^{166}\text{Ho}$ calculated MACS values are shown in Fig.~\ref{fig:MACSs} (a) and (b), respectively. 
The error propagation from the systematic and statistical errors of the NLD and GSF is done with the same procedure as in Ref.~\cite{Pogliano22}. 
For both cases we compare our results with the range covered by the TALYS models and two selected libraries used for astrophysical network calculations: JINA REACLIB~\cite{JINAREACLIB} and BRUSLIB~\cite{BRUSLIB}. 
For $^{165}\text{Ho}$, we also compare our derived data with the ones provided in the KADoNiS database~\cite{kadonis}. 
In the KADoNiS database, it is specified that the MACS measurements for this nucleus fall in two groups, one where the MACS at 30 keV is $\approx 1380$~mb~\cite{Macklin81} and one providing cross sections $\sim 15$\% lower at $\approx 1200$~mb \cite{Qaim1992,yamamuro1980,Czirr73}. 
The KADoNiS value of $1237 \pm 183$~mb is calculated as an average of the two groups, which falls slightly below (but still close to) the previous value from the compilation by Bao \textit{et al.}~\cite{Bao2000} of $1280 \pm 100$ mb.
Our derived MACS agrees quite well with the JINA REACLIB rates and the recommended KaDoNiS values, both of which fall within the MACS confidence interval, although this cannot be said for the BRUSLIB values. At 30~keV we obtained a MACS of $1494_{-272}^{+193}$~mb, slightly above the KaDoNiS recommended value, but still compatible with our findings.


For $^{166}\text{Ho}$ we show both the rates calculated using the global phenomenological OMP model from Koning \& Delaroche~\cite{localomp} and the JLM model by Bauge \textit{et al.}~\cite{JLMOMP} as these give considerably different predictions for the temperature ranges relevant to the $s$ process due to different significant contributions of the inelastic channel. 
In particular, the 30-keV MACS is estimated to be $2505^{+1257}_{-769}$~mb using the Koning \& Delaroche OMP, and $1550^{+297}_{-275}$~mb using the JLM model.

\section{Application to the \textit{s} process in AGB stars}\label{secAstrophysics}

The newly derived $^{165}\text{Ho}(n,\gamma)$ MACS may directly impact the $s$-process production of Ho. 
Assuming a local equilibrium \citep{kappeler2011}, the $A=165$ isotopic abundance $N_s(^{165}{\rm Ho})$ can be approximated by $N_s(^{165}{\rm Ho})=\langle\sigma_{164}\rangle / \langle \sigma_{165}\rangle \times N_s(^{164}{\rm Dy})$, where $\langle \sigma_{164}\rangle$ is the $^{164}$Dy MACS and $N_s(^{164}{\rm Dy})$ its $s$-process abundance. 
Therefore, a change of the $^{165}\text{Ho}(n,\gamma)$ MACS directly affects the $s$-process abundance of Ho. 
Such an impact is illustrated for the $s$ process in AGB stars, as detailed below.

AGB nucleosynthesis predictions have been computed using
the STAREVOL code \citep{Siess2008} with an
extended reaction network of 414 species linked by 637 nuclear reactions. 
Details on the nuclear network and input physics can be found in \citet{Goriely18c}.
The solar abundances are taken from \citet{asplund2009}, which correspond to a metallicity of $Z = 0.0134$. 
The \citet{Reimers1975} mass loss rate with $\eta_R=0.4$ is used from the main sequence up to the end of core helium burning and the \citet{Vassiliadis1993} prescription during the AGB phase.
Dedicated models with an initial mass of 2~$M_\odot$ and a metallicity of [Fe/H]=$-0.5$\footnote{The abundance of element X is defined as 
$\mathrm {[X/Y]} = \log_{10} (n_{\rm X}/n_{\rm Y})_*- \log_{10}(n_{\rm X}/n_{\rm Y})_\odot$ where $n_i$ is the number density of element $i$, and Y is a normalising element, generally Fe.} have been computed as explained below.

In the present calculations, a diffusion equation is used to model the partial mixing of protons in the C-rich layers at the time of the third dredge-up. 
We follow  Eq.~(9) of \citet{Goriely18c} and use the corresponding diffusive mixing parameters, {\it i.e.}, $f_{\rm env}=0.10$, $D_{\rm min} = 10^9\,{\rm cm^2\, s^{-1}}$ and $p = 5$, where $f_{\rm env}$ controls the extent of the mixing, $D_{\rm min}$ the value of the diffusion coefficient at the 
base of the envelope, and $p$ is a free parameter describing the shape of the diffusion profile. 

\begin{figure}[h]
\begin{center}
\includegraphics[width=1.1\columnwidth]{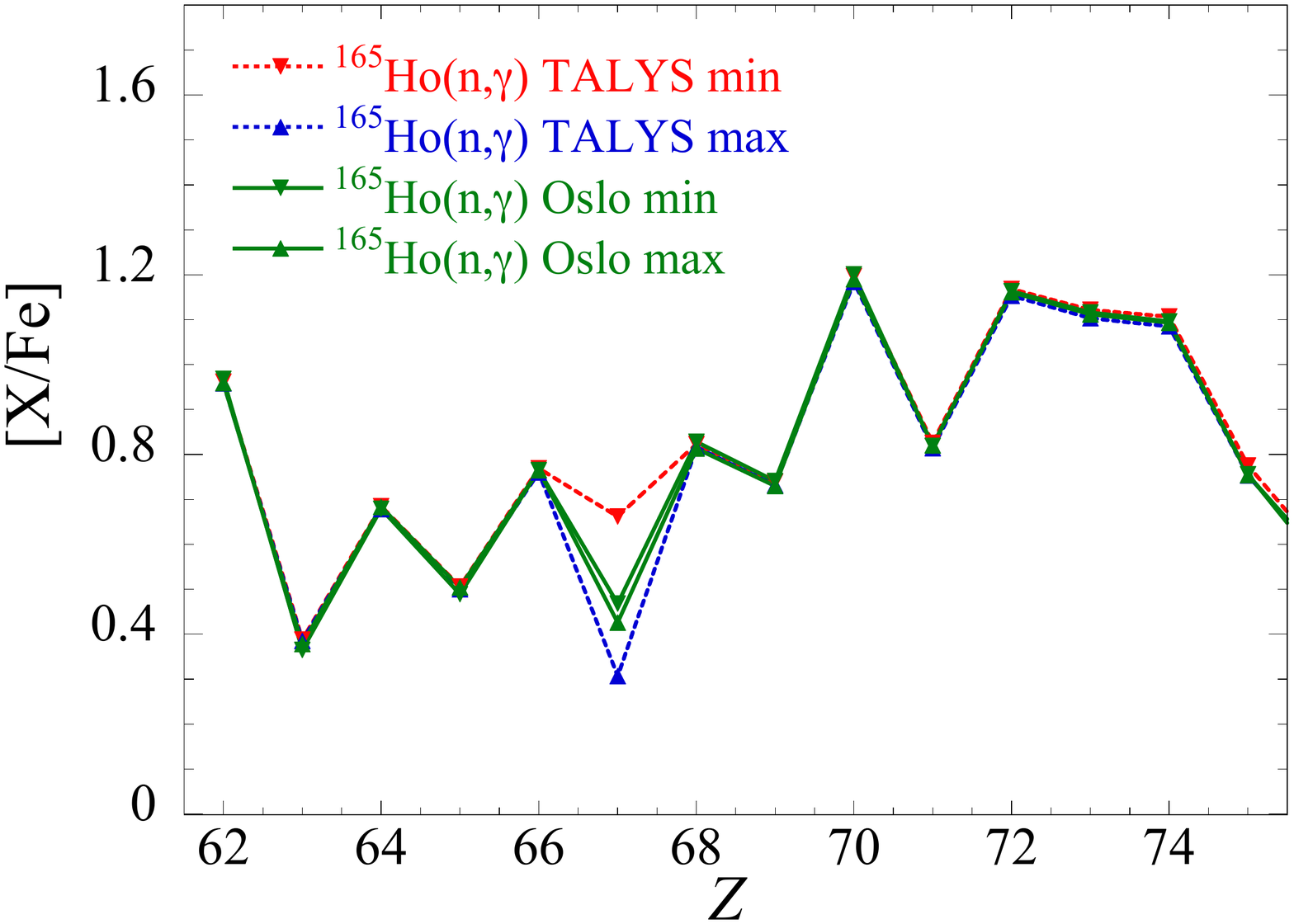}
\caption{\label{fig:XFe_Z} Elemental surface overabundances [X/Fe] at the end of the AGB evolution of our 2~$M_\odot$ [Fe/H]=$-0.5$ model star as a function of the charge number $Z$ for different values of the $^{165}\text{Ho}(n,\gamma)$ MACS, namely the upper and lower limits spanned by TALYS systematics and those constrained by the present Oslo experiment.
The Oslo-constrained $^{166}\text{Ho}(n,\gamma)$ MACS obtained with the JLM OMP is adopted in all cases.}
\end{center}
\end{figure}

The elemental surface overabundances $[{\rm X/Fe}]$ 
at the end of the AGB phase after the occurrence of 11 thermal pulses are shown in Fig.~\ref{fig:XFe_Z} for the elements ranging between Sm ($Z=62$) and Re ($Z=75$). 
On the basis of the initial large TALYS uncertainties corresponding to a variation of the $^{165}\text{Ho}(n,\gamma)$ MACS by a factor of 3.5 (see Fig.~\ref{fig:MACSs}a), an uncertainty of $\pm 0.20$~dex is obtained on the surface overabundance of Ho. 
With the newly constrained MACS, this uncertainty is reduced to $\pm 0.07$~dex. 

\begin{figure}
\includegraphics[width=1.1\columnwidth]{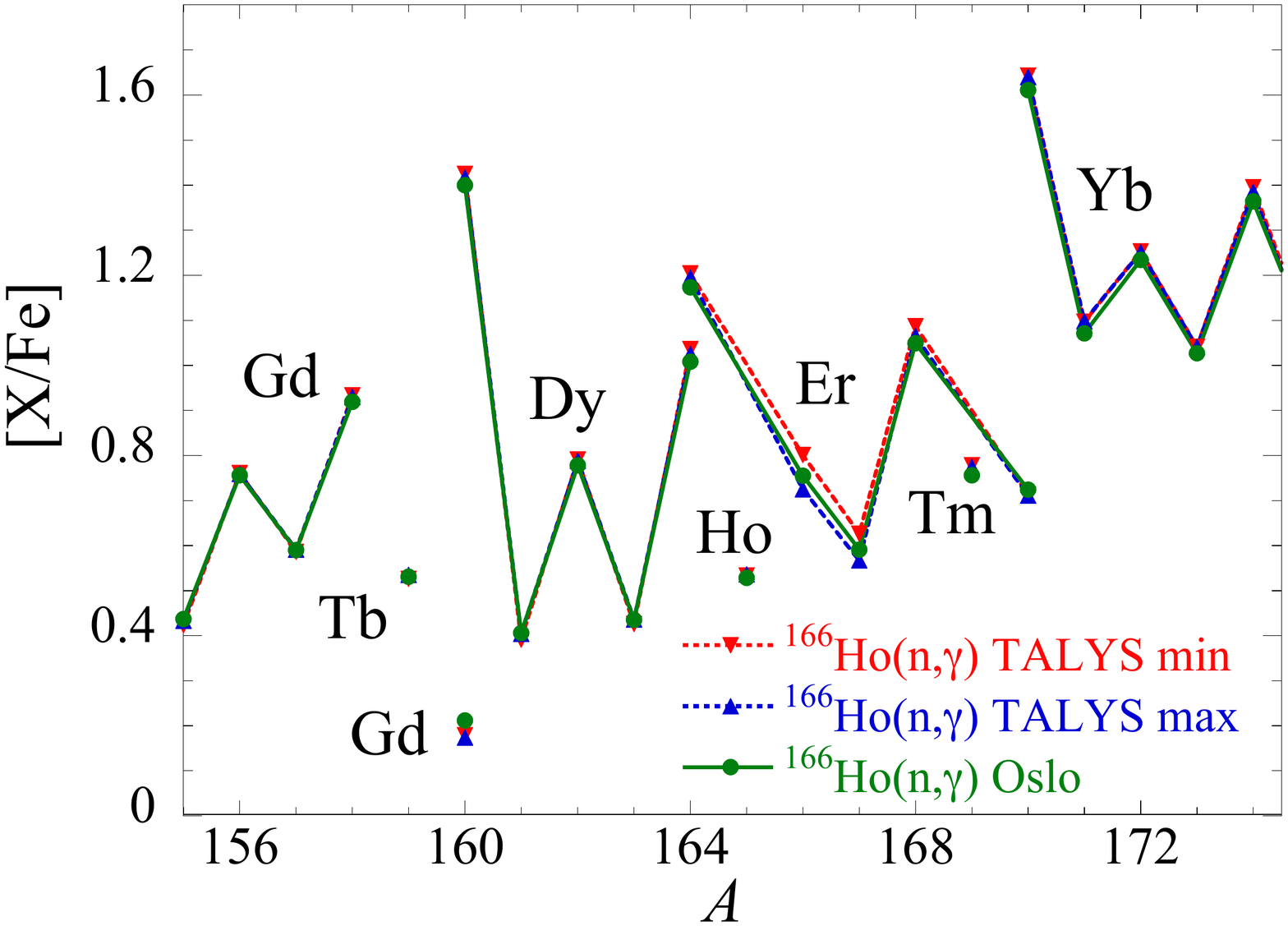}
\caption{\label{fig:XFe_A} Isotopic surface overabundances [X/Fe] at the end of the AGB evolution of our 2~$M_\odot$ [Fe/H]=$-0.5$ model star as a function of the atomic mass $A$ for different values of the $^{166}\text{Ho}(n,\gamma)$ MACS, namely the upper and lower limits spanned by TALYS systematics and those constrained by the present Oslo experiment using the JLM OMP.  }
\end{figure}

While the $^{165}\text{Ho}(n,\gamma)$ reaction directly affects the neutron capture along the radiative $s$-process path, the $^{166}\text{Ho}(n,\gamma)$ reaction comes into play only if a non-negligible amount of neutrons is produced. 
For the conditions considered here, {\it i.e.} an $s$ process during the interpulse phase at a temperature of about $10^8$~K, $^{166}$Ho can be regarded as thermalized~\cite{Misch21}, i.e its ground state and excited states are in thermal equilibrium, diminishing the potential impact of the 1200-y isomer. 
The thermalized $^{166}$Ho  half-life of $T_{1/2}\simeq 9.8$~d at $T=10^8$~K \citep{Takahashi87} is fast enough for this branching not to be affected by the $^{166}\text{Ho}(n,\gamma)$ reaction for interpulse neutron densities of $N_n \simeq 10^{7-8}{\rm cm}^{-3}$. 
Neutron densities larger than typically $3\times 10^9{\rm cm}^{-3}$ would be required for this channel to become relevant. 
Interestingly, during the convective thermal pulse, a large neutron burst may be produced by $^{22}{\rm Ne}(\alpha,n)$ and, despite a low neutron-to-seed ratio, may impact some relative isotopic abundances at the branching points. 
During the convective pulse, temperatures of $T=3-3.5\times 10^8$~K and neutron densities of $10^{10-11}{\rm cm}^{-3}$ are found. 
These latter neutron densities are high enough to activate the $^{166}\text{Ho}(n,\gamma)$ channel, despite the relatively fast $\beta$-decay of $^{166}$Ho ($T_{1/2}\simeq 1.4$~d at $T=3.5 \times 10^8$~K). 
The final isotopic surface overabundances are shown in Fig.~\ref{fig:XFe_A} for different $^{166}\text{Ho}(n,\gamma)$ rates (adopting the   
$^{165}\text{Ho}(n,\gamma)$ rate from \citet{Bao2000}). 
A lower value of $^{166}\text{Ho}$ MACS is seen to give rise to an increase of the $^{166}$Er and $^{167}$Er abundances, hence of the Er elemental overabundance by 0.04 dex, if we consider the large uncertainties spanned by TALYS calculations (see Fig.~\ref{fig:MACSs}b). 
This uncertainty is significantly reduced to below 0.01 dex, when using the Oslo-constrained rates, despite the remaining uncertainty stemming from the OMP. 
We aslo remark that if the neutron density is large enough to branch the $^{166}$Ho neutron channel, the $^{167}$Ho branching may also be slightly activated although its $\beta$-decay half-life is shorter than 3.1~h. 
This result clearly depends on the adopted TALYS rate for the $^{167}\text{Ho}(n,\gamma)$ reaction.

In summary, while the $^{165}\text{Ho}(n,\gamma)$ reaction directly affects the production of Ho, the $^{166}\text{Ho}(n,\gamma)$ reaction only plays a non-negligible role if the neutron density is high enough to activate the temperature-dependent branching at $^{166}$Ho. 
In this case, the relative enrichment of the $^{166,167}$Er isotopes may be affected. 
Through the newly derived rates, the uncertainty affecting the $s$-process abundances of Ho and Er can be significantly reduced. 
These remain much smaller than those stemming from stellar evolution modelling.

\section{Summary}
\label{secSummary}
In this work, we have presented the newly obtained NLD and GSF for rare-earth, odd-even $^{167}$Ho from the $^{164}\text{Dy}(\alpha,p\gamma)^{167}$Ho experimental data analyzed with the Oslo method. 
The NLD is shown to behave consistently to the constant-temperature model, and the GSF shows typical features for a rare-earth, neutron-rich, deformed nucleus showing structures compatible with the $M1$ scissors mode and the PDR. 
The $^{166,167}$Ho NLDs and GSFs were used to constrain the $^{165,166}\text{Ho}(n,\gamma)$ MACS uncertainties.
The MACS results were further applied to investigate the role of these two nuclei in the $s$ process.
Of particular interest is the behavior of $^{166}$Ho, whose ground state has a half-life of about 26 hours, while its 6-keV first excited state has instead a half-life of 1200 years against $\beta$ decay. 
This was studied in the context of a 2 $M_\odot$, [Fe/H]=-0.5 AGB star. 

The obtained $^{165}\text{Ho}(n,\gamma)$ MACS was shown to be lower than several of the previous experimental results, which led to a higher production of $^{165}$Ho in the $s$-process final abundances. 
With the assumption of thermalization of $^{166}$Ho in typical $s$-process interpulse conditions, the impact on the relative $^{166}$Er and $^{167}$Er enrichments is small. 
Only during convective thermal pulses were the neutron densities  high enough to activate the $^{167}$Ho branch, and consequently influence the Er abundances.

\section{Acknowledgments}
We would like to thank Pawel Sobas, Victor Modamio and Jon C. Wikne at the Oslo Cyclotron Laboratory for operating the cyclotron and providing excellent experimental conditions.
F.~P. and A.~C.~L. gratefully acknowledges funding of this research from the Research Council of Norway, project grant no. 316116. 
The calculations were performed on resources provided by Sigma2, the National Infrastructure for High Performance Computing and Data Storage in Norway (using ``Saga'' and ``Betzy'' on Project No.~NN9464K).
S.~G. and L.~S. are senior research associates from F.R.S.-FNRS (Belgium). This work was supported by the F.R.S.-FNRS under Grant No IISN 4.4502.19.
V.~W.~I., A.~G., and S.~S. gratefully acknowledge financial support from the Research Council of Norway, project number 325714. 
G.~J.~O.~F. and A.~T. acknowledge support from the INTPART program from the Research Council of Norway, project number 310094.

\bibliography{apssamp}

\end{document}